\documentclass[prl,twocolumn,showpacs,superscriptaddress,preprintnumbers,
amsmath,amssymb]{revtex4-1}

\usepackage[colorlinks=true,urlcolor=blue,linkcolor=blue,citecolor=blue]{hyperref}
\usepackage[normalem]{ulem}
\usepackage[capitalize]{cleveref}
 
\usepackage{comment}
\usepackage{flushend}
\usepackage[T1]{fontenc} 
\usepackage{mathtools}
\usepackage{xcolor}
\usepackage{graphicx}
\usepackage{dcolumn}
\usepackage{bm}
\usepackage{multirow}
\usepackage{slashed}

\usepackage{epstopdf}

\newcommand{\Beq}{\begin{equation}\begin{aligned}}
\newcommand{\Eeq}{\end{aligned}\end{equation}}

\begin{document}

\preprint{KEK-QUP-2023-0019, KEK-TH-2550, KEK-Cosmo-0321, IPMU23-0029}

\title{Regurgitated Dark Matter}

\author{TaeHun Kim}
\email{gimthcha@kias.re.kr}
\affiliation{School of Physics, Korea Institute for Advanced Study, Seoul 02455, Korea}
\affiliation{Center for Theoretical Physics, Department of Physics and Astronomy, Seoul National
University, Seoul 08826, Korea}

\author{Philip Lu}
\email{philiplu11@gmail.com}
\affiliation{Center for Theoretical Physics, Department of Physics and Astronomy, Seoul National
University, Seoul 08826, Korea}
\affiliation{International Center for Quantum-field Measurement Systems for Studies of the Universe and Particles (QUP, WPI),
High Energy Accelerator Research Organization (KEK), Oho 1-1, Tsukuba, Ibaraki 305-0801, Japan}

\author{Danny Marfatia}
\email{dmarf8@hawaii.edu}
\affiliation{Department of Physics and Astronomy, University of Hawaii at Manoa,
Honolulu, HI 96822, USA}

\author{Volodymyr Takhistov}
\email{vtakhist@post.kek.jp}
\affiliation{International Center for Quantum-field Measurement Systems for Studies of the Universe and Particles (QUP, WPI),
High Energy Accelerator Research Organization (KEK), Oho 1-1, Tsukuba, Ibaraki 305-0801, Japan}
\affiliation{Theory Center, Institute of Particle and Nuclear Studies (IPNS), High Energy Accelerator Research Organization (KEK), Tsukuba 305-0801, Japan
}
\affiliation{Graduate University for Advanced Studies (SOKENDAI), \\
1-1 Oho, Tsukuba, Ibaraki 305-0801, Japan}
\affiliation{Kavli Institute for the Physics and Mathematics of the Universe (WPI), UTIAS, \\The University of Tokyo, Kashiwa, Chiba 277-8583, Japan}

\begin{abstract}
We present a new paradigm for the production of the dark matter (DM) relic abundance
based on the evaporation of early Universe primordial black holes (PBHs) themselves formed from DM particles. As a concrete realization, we consider a minimal model of the dark sector in which a first-order phase transition results in the formation of Fermiball remnants
that collapse to PBHs, which then emit DM particles.
We show that the regurgitated DM scenario allows for DM in the mass range $\sim1$~GeV $- \,10^{16}$~GeV, thereby unlocking parameter space considered excluded.
\end{abstract}

\maketitle

{\it Introduction}.--
Dark matter (DM) constitutes $\sim85\%$ of all matter in the Universe, as determined by a multitude of astronomical observations~(for reviews, see e.g.~\cite{Bertone:2016nfn,Gelmini:2015zpa}). Despite decades of efforts to detect its non-gravitational interactions, the nature of DM remains mysterious. 

A significant focus has been on weakly interacting massive particles (WIMPs) as the DM paradigm~\cite{Steigman:1984ac, Arcadi:2017kky}, with typical masses in the GeV to multi-TeV range that often appear in theories that address the hierarchy problem of the standard model (SM).
The scenario of decoupling from the thermal bath in the early Universe, which successfully explains cosmological observations such as the light element abundances~\cite{Peebles:1991}, suggests that WIMPs with typical electroweak-scale masses and annihilation cross sections can readily account for the observed DM relic abundance through thermal freeze-out (the so-called~``WIMP miracle''). Sensitive experimental searches~(e.g.~\cite{XENON:2023cxc,LZ:2022lsv}) significantly constrain the parameter space of minimal WIMP scenarios.
DM scenarios based on additional or number-changing DM particle interactions, such as the strongly interacting massive particle miracle~\cite{Hochberg:2014dra}, provide alternative approaches to achieving the DM relic abundance.

In this work we present a novel paradigm, which we call \textit{regurgitated dark matter} (RDM), for 
producing the DM relic abundance. It is
based on the evaporation of early Universe primordial black holes (PBHs), themselves constituted by DM particles. PBHs scramble and re-emit DM particles with altered energy-momentum and abundance distributions, distinct from that of the original DM particles in the thermal bath that formed the PBHs, i.e., these properties are not determined by direct DM interactions as in conventional DM production mechanisms. While particle DM emission from evaporating PBHs has been studied (e.g.~\cite{Hooper:2019gtx,Cheek:2021odj,Marfatia:2022jiz}), in the RDM scenario, PBHs originate from the same DM particles that later constitute the DM relic abundance and not from some distinct mechanism; for PBH production mechanisms see e.g.~\cite{Zeldovich:1967,Hawking:1971ei,Carr:1974nx,GarciaBellido:1996qt,Green:2000he,Frampton:2010sw,Cotner:2019ykd,Cotner:2018vug,Green:2016xgy,Sasaki:2018dmp,Cotner:2018vug,Cotner:2019ykd,Kusenko:2020pcg,Carr:2020gox,Escriva:2022duf,Lu:2022yuc}.
As we demonstrate, RDM can open a 
new window in the
WIMP parameter space with either fermion or scalar DM particles.

{\it Model}.--
We illustrate an elegant realization of RDM in the context of a first-order phase transition (FOPT) in an asymmetric dark sector that produces Fermiball remnants composed of dark sector particles that subsequently collapse to PBHs.
The dark sector particles that form the PBHs are emitted by these PBHs through Hawking evaporation. Depending on the particle mass and PBH mass, RDM particles may be relativistic at the epoch of Big Bang nucleosynthesis (BBN) or contribute to warm DM.  We note, however, that our RDM mechanism is general and may be realized in the context of other scenarios, such as the collapse of solitonic macroscopic objects to PBHs or in models with additional dark sector forces and SM portals beyond the Higgs portal. We leave the exploration of such possibilities for future work.

We consider the model of Refs.~\cite{Hong:2020est,Kawana:2021tde,Lu:2022jnp,Kawana:2022lba,Lu:2022paj} given by
\begin{align} \label{eq:higgslang}
    \mathcal{L} =&~ {\cal L}_{\rm SM}^{} +\frac{1}{2}\partial_\mu \phi \partial^{\mu} \phi - \frac{\mu^2}{2}\phi^2 - \frac{\kappa}{2}\phi^2 (\mathcal{H}^\dagger \mathcal{H}) - V(\phi) \notag\\ 
    &~+ \Bar{\chi} i \slashed{\partial}\chi - y_\chi \phi \Bar{\chi}\chi~,
\end{align}
where ${\cal L}_{\rm SM}$ is the standard model (SM) Lagrangian, dark sector fermions $\chi$, $\Bar{\chi}$ and scalar $\phi$ interact via an attractive Yukawa force with coupling $y_{\chi}$, and $\phi$ interacts with the SM sector through the Higgs $\mathcal{H}$ doublet portal coupling $\kappa$. On receiving thermal corrections, the potential $V(\phi)$ becomes $V(\phi, T)$ which triggers a FOPT below the critical temperature, forming Fermiball remnants that collapse to PBHs. We remain agnostic about the details of the potential, allowing
a general discussion of RDM. We assume the existence of an asymmetry between the number density of $\chi$ and $\bar{\chi}$ with $\eta_{\chi} = (n_{\chi} - n_{\bar{\chi}})/s(T_\star)$, where $s(T_\star) = 2 \pi^2 g(T_\star) T_\star^3/45$ is the entropy number density with relativistic degrees of freedom (d.o.f.) $g(T)$ at temperature $T_{\star}$ of the FOPT; we neglect the small contribution of the dark sector d.o.f. to $g(T)$. 
Such an asymmetry can be realized in a variety of asymmetric DM mechanisms~\cite{Kaplan:2009ag,Petraki:2013wwa,Zurek:2013wia}, with Fermi-degenerate remnants dominated by $\chi$.

\begin{figure}[t] \centering 
\includegraphics[width=0.48\textwidth]{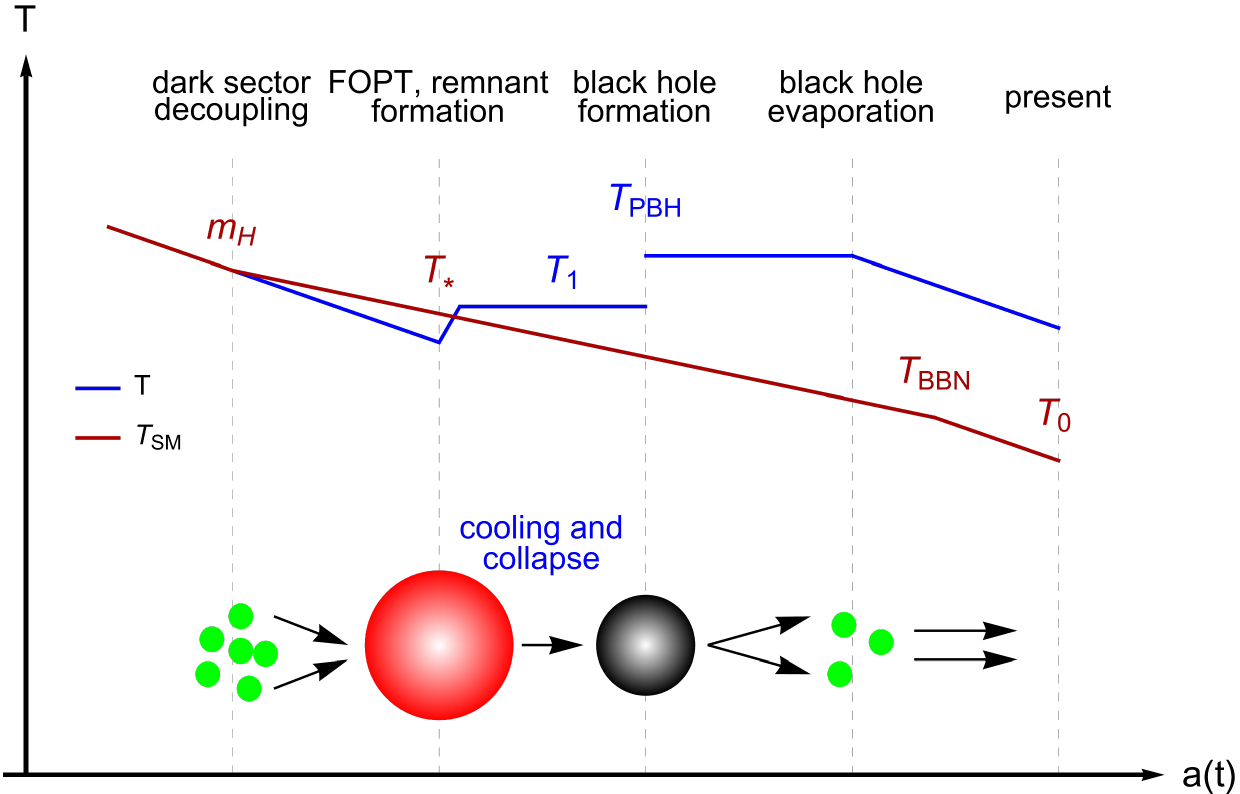}
\caption{Cosmological thermal history of RDM production. The dark sector particles in the Fermiball are re-emitted at a higher temperature after black hole formation.
} 
\label{fig:thermal history}
\end{figure}

{\it Formation of Fermiballs}.-- We consider the following thermal history of cosmology for production of RDM, illustrated schematically in Fig.~\ref{fig:thermal history}. 

At first, the dark sector and SM particles are in thermal equilibrium after inflationary reheating, which can occur either due to the Higgs portal coupling $\kappa$ or inflaton sector couplings. As the Universe expands and temperature decreases below the electroweak phase transition at $T \sim 160$~GeV, the 
dark sector decouples from the visible sector at $T \sim m_h = 125$~GeV due to the diminishing interactions between $\phi$ and the Higgs. Hereafter, the SM and dark sector temperatures $T_{\rm SM}$ and $T$, respectively, evolve separately. The effective number of relativistic degrees of freedom (d.o.f.) of the SM at dark sector decoupling is $g(T_{\rm SM}^{\rm dec})$.

At a temperature $T_\star$, a FOPT is triggered by the dark scalar potential $V(\phi,T_\star)$. The phase transition changes the expectation value of $\phi$, from $\langle \phi \rangle = 0$ in the false vacuum to $\langle \phi \rangle = v_\star$ in the true vacuum. 
The FOPT proceeds through bubble nucleation (with expanding bubble wall speed $v_{w}$) and can be characterized by the following parameters~\cite{Caprini:2019egz}: $\beta$ is the inverse duration of the FOPT, and $\alpha_D \simeq 30 \Delta V/\pi^2 g_D T_\star^4$ quantifies its strength. Here, $g_D=4.5$ is the d.o.f. of the dark sector, and $\Delta V$ is the potential energy difference between the false and true vacua.

The FOPT can readily induce a significant mass gap $\Delta m_{\phi}, \Delta m_{\chi} \gg T_\star$ between the vacua, with massless fermions acquiring mass $m_{\chi} = y_{\chi} v_\star$ via the Yukawa coupling.
For the particles to be trapped, the dark sector particle masses in the true vacuum must exceed the FOPT temperature:
\begin{equation}
m_\chi\gg T_\star,\quad m_\phi= \left(\frac{\partial^2 V(\phi,T_\star)}{\partial \phi^2}\right)^{1/2}\Big|_{\phi=v_\star}\gg T_\star^{}~.
\label{eq:trap}
\end{equation}
This condition can be fulfilled in 
supercooled scenarios with large $v_\star/T_\star$~\cite{Creminelli:2001th,Nardini:2007me,Konstandin:2011dr,Jinno:2016knw,Marzo:2018nov}, or with $y_{\chi} \gg 1$~\cite{Carena:2004ha,Angelescu:2018dkk}. Then
the mass of dark sector particles is significantly larger than their thermal kinetic energy and cannot penetrate into the true vacuum bubbles. As true vacuum bubbles expand, dark sector particles are efficiently trapped in contracting regions of false vacuum. 
In much of the parameter space, explicit calculations confirm that the  trapping fraction of dark sector particles in the false vacuum remnants is $\sim 1$ if Eq.~\eqref{eq:trap} is satisfied.
The remnants get compressed to form non-topological solitonic Fermiball remnants~\cite{Hong:2020est,Kawana:2021tde,Marfatia:2021twj,Marfatia:2021hcp,Lu:2022jnp,Kawana:2022lba,Lu:2022paj}.  We assume that the number of d.o.f. does not significantly vary between decoupling and the FOPT, which will not affect our conclusions.
 
{\it Fermiball cooling}.-- The dark sector temperature of the Fermiballs will be $T_1 = (90 \Delta V/\pi^2 g_D)^{1/4}~$, during the slow remnant cooling process~\cite{Kawana:2022lba}. 
The Fermiballs  cool via SM particle production through the Higgs portal.
A detailed analysis of Fermiball cooling for our regimes of interest can be found in Supplemental Material~\cite{supmat}. The asymmetry $\eta_{\chi}$ ensures that a population of $\chi$ particles survives after annihilation to $\phi$'s. As the Fermiball cools, these particles dominate until the Fermiball collapses into a black hole.

The dominant cooling channel for Fermiballs depends on the dark sector temperature $T_1$.
For $T_1$
below the electroweak scale, the cooling rate is suppressed by the Higgs mass. 
For small values of $\kappa$, we have verified that volumetric cooling occurs with a rate $\dot{C} = n^2 \langle 2E \rangle \sigma v_{\rm rel}$ (see Supplemental Material~\cite{supmat} for details). The scalar $\phi$ follows a thermal Bose-Einstein distribution with number density $n=(\zeta(3)/\pi^2) T^3$ and average energy $\langle E \rangle \simeq 2.7 T_1$.

In this regime, 
Fermiball remnants will predominantly cool 
through $\phi\phi \xrightarrow{} f\Bar{f}$ emission, where $f$ is the heaviest available fermion with mass $m_f$.
The remnant is initially supported by thermal pressure, but as it cools, it becomes supported by the Fermi degeneracy pressure of the asymmetric $\chi$ population. In the volumetric cooling regime, this transition happens at a temperature,
\begin{align}
\begin{split}
\label{eq:ttrlow}
     T_{\rm SM}^{\rm tr, low} \simeq&~ (10^4~\textrm{GeV})\, \kappa \left(\frac{T_1}{1~\textrm{GeV}}\right)^{3/2} \frac{m_f}{1.27\textrm{ GeV}} \\ & \times\left(\frac{g_D}{4.5}\right)^{-1/2}   \left(\frac{g(T_{\rm SM}^{\rm tr})}{106.75}\right)^{-1/4}
     ~.
\end{split}
\end{align}
For $T_1$ 
above the electroweak scale, direct Higgs production, $\phi \phi \rightarrow \mathcal{H} \mathcal{H}$, can occur.  
For small $\kappa$, we have volumetric cooling with the transition temperature,
\begin{align}
\begin{split}
\label{eq:ttrhigh}
    T_{\rm SM}^{\rm tr, high} \simeq &~ (10^4\textrm{ TeV})\, \kappa \left(\frac{T_1}{1\textrm{ TeV}}\right)^{1/2} \left(\frac{g_D}{4.5}\right)^{-1/2}~.
\end{split}
\end{align}
For larger $\kappa$ in both temperature ranges, the mean free path of the particles can become shorter than the Fermiball radius, and blackbody surface cooling dominates. In this case, the transition time is negligible, i.e., $T_{\rm SM}^{\rm tr}\sim T_\star$ (see Supplemental Material~\cite{supmat} for details). Hence, the resulting DM abundance is determined primarily by the black hole evaporation timescale.

{\it Black hole formation}.--  Black holes are formed from the cooling Fermiballs
when the length scale associated with attractive Yukawa force $\sim1/m_{\phi}$ is comparable with the mean separation of $\chi$ ($\overline{\chi}$) inside. In practice, this collapse occurs shortly after the transition to Fermi degeneracy pressure, at temperature $T_{\rm SM}^{\rm tr}$~\cite{Kawana:2022lba,Lu:2022paj}. (Note that gravitational collapse is also possible~\cite{Gross:2021qgx}.)
This instability is ensured for $\alpha_D>0.01$~\cite{Lu:2022paj}, which is readily satisfied for $\alpha_D >1/3$ so that the initial remnant shrinks efficiently.
Then, the average mass of PBHs formed from the collapse of Fermiballs is~\cite{Kawana:2022lba}
\begin{align}
\begin{split}
    \label{eq:mass}
    \overline{M}_{\rm PBH}^{} \simeq~ & (6.61\times10^{14}\, {\rm g})~   \alpha_D^{1/4} \left(\frac{v_w}{0.7}\right)^3 
    \frac{\eta_\chi}{10^{-10}} \left(\frac{\beta/H}{1000}\right)^{-3}\\ & \times  \left(\frac{g(T_{\star})^{}}{g(T_{\rm SM}^{\rm dec})}\right)^{-2/3}
    \left(\frac{T_{\star}^{}}{1~\rm{GeV}}\right)^{-2}~. 
\end{split}
\end{align}
The number density of these PBHs
is
\begin{align}
\begin{split}
\label{eq:nT}
    n_{\rm PBH}(T_{\rm SM}) =~ & 3.83\times10^8 \left(\frac{v_w}{0.7}\right)^{-3}\left(\frac{\beta/H}{1000}\right)^3 \\ & \times \frac{g_s(T_{\rm SM})g(T_\star)^{1/2}T_{\rm SM}^3 T_\star^3}{M_{\rm pl}^3}
\end{split}
\end{align}
with $g_s(T_{\rm SM})$ the number of entropic d.o.f. of the visible sector.

PBHs will eventually evaporate through Hawking radiation~\cite{Hawking:1974rv} at a time $t_{\rm PBH}$ after formation, as reviewed in Supplemental Material~\cite{supmat}. If PBHs come to dominate the matter density, then evaporation will reheat the Universe to a temperature,
\begin{equation}
\label{eq:trhm}
    T_{\rm SM}^{\rm RH} = 50.5 \textrm{ MeV} \left[\frac{ \overline{M}_{\rm PBH}^{} }{10^{8}~{\rm g}}\right]^{- {3 \over 2}} \left[\frac{g(T_{\rm SM}^{\rm RH})}{10}\right]^{-{ 1 \over 4}}\left[\frac{g_{\rm H,SM}}{108}\right]^{1 \over 2},
\end{equation}
where $g_{\rm H,SM}$ is the number of Hawking d.o.f. for the SM sector, which we elaborate on below.
If there is no PBH-dominated era, then the prefactor in Eq.~\eqref{eq:trhm} becomes $T_{\rm SM}^{\rm evap}\sim 43\textrm{ MeV}$. The time at which evaporation occurs is the sum of the three timescales -- the PBH lifetime, the formation (cooling) time, and $t_\star$. Because of the hierarchy of scales, the evaporation occurs at $\sim \min(T_{\rm SM}^{\rm tr}, T_{\rm SM}^{\rm RH},T_\star$).

There are several key timescales in our setup.  We are primarily interested in PBH evaporating into sufficient quantities of RDM before BBN to maintain the successful predictions of the light element abundances.
This sequence includes the production of Fermiballs, 
their subsequent cooling and collapse into PBHs, and PBH evaporation. The relevant timescales for Fermiball formation and PBH collapse are related to the temperatures $T_{\star}$ and $T_{\rm SM}^{\rm tr}$ respectively, and the evaporation temperature is $T_{\rm SM}^{\rm RH}$ ($T_{\rm SM}^{\rm evap})$.
To evade BBN constraints, we require that $T_{\star}, T_{\rm SM}^{\rm tr}, T_{\rm SM}^{\rm RH} \geq 5\textrm{ MeV}$.
The condition for PBH domination before they evaporate is 
$ \overline{M}_{\rm PBH} n_{\rm PBH}(T_{\rm SM}^{RH}) > \rho_{\rm SM}$.

\begin{figure*}[t] \centering 
\includegraphics[width=0.48\textwidth]
{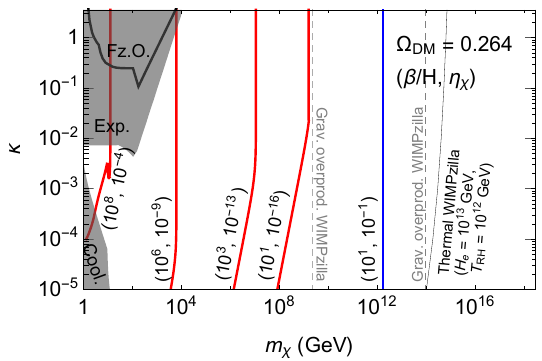}
\includegraphics[width=0.48\textwidth]
{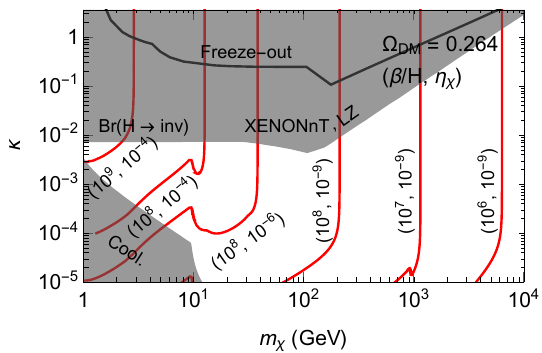}\caption{
Constraints on the Higgs portal coupling $\kappa$ as a function of $m_{\chi}$ (assuming $m_\phi = m_\chi$) over a wide mass range [left] and for the WIMP mass range [right]. On the solid iso-$\Omega_{\rm DM}=0.264$ contours, $\chi$ constitutes the entire DM abundance for specific choices of the inverse duration of the FOPT $\beta/H$ and dark sector fermion asymmetry $\eta_{\chi}$. 
The PBHs dominate the Universe's energy density on the blue contours and are subdominant on the red contours.
We fix $\alpha_D = 1$, $v_{w} = 0.67$, $g_{D, \ast} = 4$
and $T_1 \simeq T_{\star} \simeq 0.05 m_\chi$. 
Also displayed are the most restrictive bounds from several experiments (labeled "Exp." in the left panel),  including the 95\% CL bound from
the invisible Higgs decay branching fraction~\cite{Arcadi:2019lka}, and the 90\% CL bounds from XENONnT~\cite{XENON:2023cxc} and LUX-ZEPLIN (LZ)~\cite{LUX-ZEPLIN:2022xrq}. The solid black curve corresponds to conventional Higgs portal WIMP freeze-out,  and the solid and dashed gray curves correspond to thermal production and gravitational overproduction of superheavy WIMPZillas, respectively. In the lower shaded region labeled "Cool.", PBHs form after BBN. 
} 
\label{fig:exclusionsigmaDN}
\end{figure*}

{\it Regurgitated dark matter}.--  The dark sector particles are regurgitated during PBH evaporation and can constitute the main component of DM either in the WIMP mass range, $1~\textrm{GeV}\lesssim m_{\chi} \lesssim 1$~TeV, or the superheavy mass range with $m_{\chi} \gg 1$~TeV.
The energy density of RDM depends on the PBH mass fraction at evaporation. If the PBHs evaporate during the radiation dominated era, the density of regurgitated particles is suppressed by $\rho_{\rm PBH}/(\rho_{\rm PBH}+\rho_{\rm SM})\simeq \rho_{\rm PBH}/\rho_{\rm SM}$. The ratio is unity if PBHs dominate the matter density.

The relative Hawking emission rates of dark sector particles to SM particles is given by the ratio of their effective Hawking d.o.f. $g_{\rm H,D}/g_{\rm H,SM}$, 
with $g_{\rm H,D} = 5.82$ for the dark sector and $g_{\rm H,SM}=108$ for the SM sector~\cite{Page:1976df, MacGibbon:1990zk} (see Supplemental Material~\cite{supmat}). Most emitted particles have masses smaller than the Hawking temperature $T_{\rm PBH}$ of the PBH. 
So, $T_{\rm PBH} \gtrsim m_\chi,m_\phi \gg T_1, T_\star$ holds for dark sector particle emission.
While both $\phi$ and $\chi$ are emitted, the scalar $\phi$ develops a mixing with the SM Higgs after the electroweak and dark sector phase transitions, allowing for the decay of $\phi$ into SM particles. In the relevant regions of allowed parameter space, this decay timescale is shorter than the lifetime of the Universe. Hence,
in this particular model realization of the RDM paradigm, $\phi$ does not constitute a significant DM component. In the following, we focus on the abundance of the stable fermion $\chi$. Although $\phi$ is unstable and does not contribute to the current mass density, it's parameters can still be constrained by considerations of the cooling rate and the Higgs invisible decay. In these instances, we assume $m_\phi = m_\chi$.  

When the initial Hawking temperature is smaller than the particle mass, the dark sector particles are emitted once the Hawking temperature reaches their mass in the true vacuum,  
$\epsilon_{\rm em} T_{\rm PBH} \simeq m_{\chi}$ where $\epsilon_{\rm em} = 2.66$, $4.53$, $6.04$ for spin $s = 0, 1/2, 1$ particles, respectively~\cite{MacGibbon:1991tj}. This higher temperature corresponds to a lighter PBH mass below which these heavy particles can be emitted:
\begin{equation}
\label{eq:Mem}
    M_{\rm PBH}^{\rm em} = (1.06\times10^8\,  {\rm g}) \, \epsilon_{\rm em} \left(\frac{m_{\chi}}{10^5\textrm{ GeV}}\right)^{-1}~.
\end{equation}
The bulk of emitted particles will be nonrelativistic, with the initial density at emission reduced by a factor of $(M_{\rm PBH}^{\rm em}/\overline{M}_{\rm PBH})$, such that
$\rho_{\chi}/\rho_{\rm SM} = (M_{\rm PBH}^{\rm em}/\overline{M}_{\rm PBH}) (g_{\rm H,\chi}/g_{\rm H,SM})$.

From the emission spectrum of PBHs, the average energy for heavy particles is $\overline{E} = 2 m_{\chi}$. Thus, the present day DM mass density of such primarily 
nonrelativistic emitted 
DM particles is 
\begin{align}
\begin{split}
    \Omega_{\rm DM,\chi}^{\rm NR} =&~ \frac{\Omega_{r,0}}{2} \frac{g_{s}^{1/3}(T_{\rm SM}^{\rm RH}) T_{\rm SM}^{\rm RH}}{g_{s}^{1/3}(T_{\rm SM}^{0}) T_{\rm SM}^{0}} \frac{M_{\rm PBH}^{\rm em}}{\overline{M}_{\rm PBH}} \frac{g_{\rm H,\chi}}{g_{\rm H,SM}} \\ 
    =&~ 1.24\times10^{5}\, \epsilon_{\rm em} \, g_{H,\chi} \left(\frac{m_{(\phi,\chi)}}{10^5\textrm{ GeV}}\right)^{-1} \\
    &\times\left(\frac{\overline{M}_{\rm PBH}}{10^8 ~{\rm g}}\right)^{-5/2} \left(\frac{g(T_{\rm SM}^{\rm RH})}{10}\right)^{1/12}
\end{split}
\end{align}
where $T_{\rm SM}^{0} =2.73$~K and $g_s(T_{\rm SM}^{0})=3.9$ are the present day values.  An additional factor of $\rho_{\rm PBH}/\rho_{\rm SM}$ appears if the PBHs are subdominant (see Supplemental Material~\cite{supmat}).

If the particles are emitted relativistically, the present day mass density contribution is reduced by the redshifting of these particles until they become 
nonrelativistic. The bulk of the Hawking radiation emitted particles will have a Lorentz boost factor $\gamma \simeq \epsilon_{\rm em} T_{\rm PBH}/m_{\chi}$.
The resulting density of these initially relativistic dark sector particles is 
\begin{equation} 
\label{eq:omegachirel}
     \Omega_{\rm DM,\chi}^{\rm R} = \Omega_{\rm DM,\chi}^{\rm NR} \times \frac{4}{3}
     \frac{m_\chi}{\epsilon_{\rm em} T_{\rm PBH}} \frac{\overline{M}_{\rm PBH}}{M_{\rm PBH}^{\rm em}}~.
\end{equation}
The behavior is opposite to the nonrelativistic case, with lighter particles being less dense.

\textit{Dark matter detection.}--
As dark matter, $\chi$ can be observed in direct detection experiments by interactions through the Higgs portal coupling in Eq.~\eqref{eq:higgslang}. The resulting elastic scattering cross section of $\chi$ on nucleons $N$ is given by~\cite{Arcadi:2019lka, Arcadi:2021mag} (see Supplemental Material~\cite{supmat} for details)
\begin{align}
\begin{split}
    \sigma_{\chi N} = &~ \frac{\kappa^2}{\pi m_h^4} \left(\frac{m_\chi^2}{m_\phi^2 - m_h^2} \right)^2 \frac{m_N^4 f_N^2}{(m_\chi + m_N)^2} 
    \\ \sim &~ \frac{3.5\times 10^{-38} \, \text{cm}^2 \times \kappa^2}{\left(\dfrac{m_\chi}{1~\text{GeV}} + 1\right)^2}\times  \left(\frac{m_\chi^2}{m_\phi^2 - m_h^2} \right)^2~,
    \label{eq:HiggsportalsigmachiN}
\end{split}
\end{align}
where $m_N \simeq 1 \, \text{GeV}$ is the nucleon mass and $f_N \sim 0.3$ is Higgs-nucleon interaction parameter.

Employing Eq.~\eqref{eq:HiggsportalsigmachiN}, we recast existing bounds on the spin-independent scattering cross section into constraints on the coupling $\kappa$.
In Fig.~\ref{fig:exclusionsigmaDN}, we display constraints on $\kappa$ (upper shaded region)  as well as predictions (solid colored lines) for $\chi$ to constitute all of the DM abundance for different values of $\beta/H$ and $\eta_{\chi}$. In the lower shaded region labeled ``Cool.'',  PBHs form after BBN via Fermiball cooling. Clearly, the regurgitated $\chi$ can saturate the DM relic abundance for a wide range of masses. The final abundance has a $\kappa$ dependence if the cooling time is longer than the lifetime of the PBH and the transition time $t_\star$. In the WIMP mass range, the cooling rate depends on fermion channels that open sequentially with increasing $m_\chi=20 T_1$. Here, $g(T_\star)$ also changes similarly and affects the PBH number density through Eq.~\eqref{eq:nT}. 

The Yukawa interaction between $\chi$ and $\phi$ keeps $\chi$ in thermal equilibrium after it decouples from the SM. Therefore, we calculate $\chi$'s freeze-out abundance by assuming the two particles freeze-out together when $\phi$ decouples from the SM plasma. We interpret the results for thermal freeze-out production of Ref.~\cite{Steigman:2012nb} for $\phi \phi \rightarrow f \overline{f}$ in the nonrelativistic limit, (see Supplemental Material~\cite{supmat}), and plot the result as the solid black curve in  Fig.~\ref{fig:exclusionsigmaDN}. While WIMP masses are strongly constrained, RDM can be efficiently produced in the unconstrained parameter space below $m_{\chi}\sim$ TeV.

For $m_\chi < m_h / 2$, stringent bounds on $\kappa$ arise from the invisible Higgs decay branching fraction to DM particles Br$(H \rightarrow {\rm inv})$ constrained by Large Hadron Collider (LHC) data at 95\% confidence level~\cite{Arcadi:2019lka}. We show this bound in Fig.~\ref{fig:exclusionsigmaDN} with $m_\phi = m_\chi$.
Direct detection constraints on $\kappa$ weaken for heavier $m_\chi$ since $\sigma_{\chi N} \propto 1/m_\chi^2$ for $m_\chi \sim m_\phi \gg m_h \gg m_N$.  

In the mass range $10^{9} \, \text{GeV} \lesssim m_{\chi} \lesssim 10^{14} \, \text{GeV}$, gravitational overproduction of WIMPZillas~\cite{Fedderke:2014ura, Chung:1998zb, Kuzmin:1998kk, Kolb:1998ki} can restrict $\chi$ from being a viable DM candidate depending on the Hubble rate $H_e$ at the end of inflation~\cite{Kolb:2017jvz}; see the gray dashed lines in Fig.~\ref{fig:exclusionsigmaDN}. However, if inflation occurs at a lower energy scale the abundance of WIMPZillas can be significantly suppressed.

If the WIMPZilla has additional interactions, a thermal relic can be realized. For the Higgs portal scenario~\cite{Kolb:2017jvz}, we show the case of $H_e = 10^{13} \, \text{GeV}$ with the reheating temperature $T_{\rm SM}^{\rm RH} = 10^{12} \, \text{GeV}$.
For even heavier DM masses, $m_{\chi} \gtrsim 3 \times 10^{16} \, \text{GeV}$, stringent bounds originate from DEAP-3600 and mica searches~\cite{Carney:2022gse}, but they fall in a region with no reliable scaling relation. 
Furthermore, a robust theoretical bound for pointlike DM~\cite{Digman:2019wdm} cannot be meaningfully applied to $\kappa$, since this would imply that the nuclear scattering cross section of $\chi$ plateaus to a maximum value even for an arbitrarily large $\kappa$.

Gravitational waves (GWs) from the FOPT can provide a correlated signature with DM detection in scattering experiments.
While detailed predictions depend on specifics of the FOPT, the peak GW frequency is expected to be
$f_{\rm GW} \simeq \mathcal{O}(10^{-2})\textrm{mHz}(\beta/H_{\star}/100)(T_{\star}/1~\textrm{GeV})$,
which could fall in the range of upcoming interferometers such as
LISA~\cite{amaroseoane2017laser}, Einstein Telescope~\cite{Punturo:2010zz},
Cosmic Explorer~\cite{Reitze:2019iox}, Big Bang Observer~\cite{Crowder:2005nr} and DECIGO~\cite{Seto:2001qf}. Moreover, if PBHs dominate the matter density,  their evaporation can lead to induced GWs~\cite{Inomata:2019ivs,Inomata:2020lmk,Domenech:2020ssp,Domenech:2021wkk,Domenech:2021ztg}.
We leave the study of the associated GW signal for future work.
 
\textit{Conclusions.}--- We proposed a novel paradigm of regurgitated DM, stemming from the emission of evaporating PBHs formed from the DM particles themselves. This is distinct from conventional particle DM production mechanisms, since the resulting DM relic abundance is not determined by particle interactions. Intriguingly, as we demonstrate with a concrete realization, this paradigm can produce the inferred abundance of DM in a very broad mass range $\sim 1$~GeV $-\, 10^{16}$~GeV, and opens up parameter space previously thought to be excluded.
A stochastic background of GWs is a possible correlated signature of the scenario. 

~\newline
\textit{Acknowledgements.} We thank J.~Arakawa, J.~Jeong, S.~Jung, K.~Kawana, J.~Kim and K. Xie for useful discussions. T.H.K. is supported by a KIAS Individual Grant No. PG095201 at Korea Institute for Advanced Study and National Research Foundation of Korea under Grant No. NRF-2019R1C1C1010050. P.L. is supported by Grant Korea NRF2019R1C1C1010050. D.M. is supported in part by the U.S. Department of Energy under Grant No. DE-SC0010504.
V.T. acknowledges support by the World Premier International Research Center Initiative (WPI), MEXT, Japan and JSPS KAKENHI grant No. 23K13109. This work was performed in part at the Aspen Center for Physics, which is supported
by the National Science Foundation grant PHY-2210452.

 
\bibliography{references}

\clearpage
\onecolumngrid
\begin{center}
   \textbf{\large SUPPLEMENTAL MATERIAL \\[.1cm] Regurgitated Dark Matter}\\[.2cm]
  \vspace{0.05in}
  {TaeHun Kim, Philip Lu, Danny Marfatia, Volodymyr Takhistov}
\end{center}

\twocolumngrid
\setcounter{equation}{0}
\setcounter{figure}{0}
\setcounter{table}{0}
\setcounter{section}{0}
\setcounter{page}{1}
\makeatletter
\renewcommand{\theequation}{S\arabic{equation}}
\renewcommand{\thefigure}{S\arabic{figure}}
\renewcommand{\thetable}{S\arabic{table}}

\onecolumngrid

We provide additional details of RDM production. Specifically, we discuss Hawking evaporation, Fermiball cooling rates, the conditions for PBH domination, and the interactions after the FOPT.
 
\section*{PBH Evaporation}

The lifetime of a PBH depends sensitively on $M_{\rm PBH}$ as $t_{\rm PBH}\propto M_{\rm PBH}^3$, with a PBH of initial mass $\sim 4\times10^{8}$~g evaporating around the time of
BBN. The Hawking temperature of a PBH is~\cite{Hawking:1974rv}
\begin{equation}
\label{eq:tbh}
    T_{\rm PBH} = 1.06\times10^{5}\textrm{ GeV} \left(\frac{M_{\rm PBH}}{10^{8} {\rm g}}\right)^{-1}~.
\end{equation}
The black hole emission further depends on the spin of the particles, with the relative rates of emission per d.o.f. with respect to spin-1/2 fermions given by~\cite{Page:1976df,MacGibbon:1990zk}
\begin{equation}
\label{eq:gh}
 g_H = \sum_i w_i g_{H,i}~,
~~~g_{H,i}=\left\{
\begin{aligned}
    &=  1.82,~s=0 \\
    &=  1.0,~~s=1/2 \\
    &=  0.41,~s=1 \\
    &=  0.05,~s=2
\end{aligned}
\right.
\end{equation}
where $w_i$ is the number of spin states of each particle.
The total emission rate is
\begin{equation}
\label{eq:emrate}
    \frac{dM_{\rm PBH}}{dt} = -(7.6\times10^{24}~{\rm g}/{\rm s})~g_{H,i}(T_{\rm PBH})\left(\frac{M_{\rm PBH}}{1~{\rm g}}\right)^{-2}~.
\end{equation}
The SM d.o.f. contribute a total of $g_H \simeq 108$. Using Eq.~\eqref{eq:gh}, the dark sector fermions (four spin-1/2 d.o.f.) and scalar (one spin-0 d.o.f.) contribute $\simeq 5.82$, resulting in dark sector emission contributing approximately $\sim 5.1\%$ of the total emission once the PBH temperature becomes sufficiently high for efficient emission of the dark sector particles.

From Eq.~\eqref{eq:tbh}, the shape of the integrated emission spectrum can be obtained. Differentiating, $dM_{\rm PBH}/dT_{\rm PBH}\propto T_{\rm PBH}^{-2}$ so that $E dN/dE \propto T_{\rm PBH}^{-2}$ or $dN/dE \propto T_{\rm PBH}^{-3}$ for the high energy tail. The scaling relations for the emitted particle number density are given by~\cite{MacGibbon:1991tj}
\begin{equation}
    E \gtrsim T_{\rm PBH}(M_i): \frac{dN}{dE} \propto E^{-3},~
~~~ ~E\ll T_{\rm PBH}(M_i): \frac{dN}{dE}=\left\{
\begin{aligned}
    &=  E,~~s=0 \\
    &=  E^2,~s=1/2 \\
    &=  E^3,~s=1
\end{aligned}
\right.
\end{equation}
with $M_i$ being the initial mass of the PBH. We can construct an approximate integrated spectrum by connecting the two scaling relations at the (instantaneous) peak energy $E=\epsilon_{\rm em} T_{\rm PBH}$.
The spectrum can be normalized to the PBH mass, as each primary DM particle (before decay) should represent $g_{H,D}/(108+g_{H,D})$ of the total emission. For dark fermions, we use $\epsilon_{\rm em}=4.53$ and $g_{H,D}=5.82$. Then,
\begin{equation}
\label{eq:dist}
    E \frac{dN}{dE} = \frac{g_{H,D}}{108+g_{H,D}}\frac{4M_{\rm PBH}}{5\epsilon_{\rm em} T_{\rm PBH}} \times 
 \left\{\begin{aligned}
    &\left(\frac{E}{\epsilon_{\rm em} T_{\rm PBH}}\right)^3,~~~E<\epsilon_{\rm em} T_{\rm PBH} \\
    &\left(\frac{E}{\epsilon_{\rm em} T_{\rm PBH}}\right)^{-2},~E\geq \epsilon_{\rm em} T_{\rm PBH}
\end{aligned}
\right.
\end{equation}
We make the common assumption that the emission takes place nearly instantaneously on cosmological time scales.

\section{Fermiball Cooling} 

Here we provide additional details for Fermiball cooling through the Higgs portal.
For energies and masses below the electroweak scale, the cross section for $\phi\phi \xrightarrow[]{} f\Bar{f}$ through a Higgs propagator  is
\begin{equation}
    \sigma = \frac{2\kappa^2 m_f^2 (s-4m_f^2)^{3/2}}{s^{3/2} \pi v_{\rm rel} (s-m_h^2)^2}~,
\end{equation}
where $\sqrt s$ is the center of mass energy. 
The cross section is dominated by the heaviest fermion kinematically allowed. With the hierarchy $m_f \ll m_\phi \ll m_h$, we find
\begin{equation}
    \sigma = \frac{2\kappa^2 m_f^2}{\pi v_{\rm rel} m_h^4}~.
\end{equation}
In our freeze-out calculations, we take the limit of heavy nonrelativistic $\phi$.

When $T_1$ is below the electroweak scale, the cooling proceeds primarily through volumetric cooling~\cite{Kawana:2021tde,Kawana:2022lba} with the rate
\begin{equation}
\label{eq:coollow}
    \dot{C} = n^2 \langle 2E \rangle \sigma v_{\rm rel} = \frac{0.051 \kappa^2 T_1^7 m_f^2}{m_h^4}~,
\end{equation}
where $n=(\zeta(3)/\pi^2)T^3$ and $\langle E \rangle = 2.7T_1$ for scalars. The transition temperature is then~\cite{Kawana:2022lba}
\begin{equation}
\label{eq:ttrlowapp}
    T_{\rm SM}^{\rm tr} = \left(\frac{a\dot{C}}{6\rho_d \ln(R_1/R_{\rm tr})}\right)^{1/2} = (4313\textrm{ GeV})\, \kappa \left(\frac{T_1}{\textrm{ GeV}}\right)^{3/2} \frac{m_f}{1.27~\textrm{GeV}} \left(\frac{g_D}{4.5}\right)^{-1/2}\left(\frac{g(T_{\rm SM}^{\rm tr})}{106.75}\right)^{-1/4} \left(\ln\Big(\frac{R_1}{R_{\rm tr}}\Big)\right)^{-1/2}~.
\end{equation}
Here, $\ln(R_1/R_{\rm tr})^{-1/2}$ is an $\mathcal{O}(1)$ factor that depends on the ratio of the initial Fermiball radius $R_1$ to the transition time radius $R_{\rm tr}$ approximately as $R_1/R_{\rm tr} \sim\eta_\chi^{-1/3}$.

For the case when $T_1$ is above the electroweak scale, we have direct Higgs production through $\phi \phi \xrightarrow[]{} HH$ with cross section
\begin{equation}
    \sigma = \frac{\kappa^2}{8\pi v_{\rm rel} s} \sim \frac{\kappa^2}{32\pi v_{\rm rel} m_\phi^2}~.
\end{equation}
The corresponding cooling rate can be calculated as
\begin{equation}
\label{eq:coolhigh}
    \dot{C} = 2.73\times10^{-5} \kappa^2 T_1^5~,
\end{equation}
which yields the corresponding transition temperature,
\begin{align}
\begin{split}
\label{eq:ttrhighapp}
    T_{\rm SM}^{\rm tr} = (4.05\times10^4\textrm{ TeV})\, \kappa \left(\frac{T_1}{1\textrm{ TeV}}\right)^{1/2} \left(\frac{g_D}{4.5}\right)^{-1/2}\left(\ln\Big(\frac{R_1}{R_{\rm tr}}\Big)\right)^{-1/2}~.
\end{split}
\end{align}
However, for large $\kappa$ the mean free path of the Higgs becomes shorter than the size of the Fermiball. Surface cooling happens when $n\sigma R_1 \sim \mathcal{O}(1)$. Above the electroweak scale, this occurs for
\begin{equation}
    \kappa \gtrsim 2\times10^{-4} \left(\frac{v_w}{0.7}\right)^{-1/2} \left(\frac{T_1}{1\textrm{ TeV}}\right)^{1/2}\left(\frac{\beta/H}{1000}\right)^{1/2}\left(\frac{4\alpha_D}{1+\alpha_D}\right)^{1/6}\left(\ln\Big(\frac{R_1}{R_{\rm tr}}\Big)\right)^{-1/2}~.
\end{equation}
The cooling approaches blackbody radiation so that the transition temperature becomes~\cite{Lu:2022paj}
\begin{equation}
    T_{\rm SM}^{\rm tr} \simeq (7.29\textrm{ TeV}) \left(\frac{v_w}{0.7}\right)^{-1/2}\left(\frac{T_1}{1~\textrm{TeV}}\right) \left(\frac{\beta/H}{1000}\right)^{1/2}\left(\frac{4\alpha_D}{1+\alpha_D}\right)^{1/6}~.
\end{equation}
This can also occur below the electroweak scale for sufficiently large $\kappa$. Under these approximations, the transition temperature can seem to be higher than the phase transition temperature. In practice, this means the cooling is very rapid and the transition happens within a Hubble time. The PBH is quickly formed, and the transition temperature is $\sim T_\star$.

Now we compare the cooling timescales with the evaporation timescales. From Eqs.~\eqref{eq:evap} and~\eqref{eq:ttrlowapp}, below the electroweak scale, the condition for the formation timescale  to be longer than the PBH lifetime is
\begin{equation}
\label{eq:formlowcond}
    \kappa < 6.06\times10^{-6} \left(\frac{M_{\rm PBH}}{10^8 g}\right)^{-3/2} \left(\frac{T_1}{\textrm{GeV}}\right)^{-3/2}\left(\frac{m_f}{1.27\textrm{ GeV}}\right)^{-1}\left(\ln\Big(\frac{R_1}{R_{\rm tr}}\Big)\right)^{-1/2}~.
\end{equation}
 If this condition is satisfied, the PBH density at evaporation depends on the cooling time, and is given by 
\begin{align}
\begin{split}
\label{eq:evapfracformlow}
    \frac{\rho_{\rm PBH}}{\rho_{\rm SM}}\bigg|_{\rm form} =&~ 7.58\times 10^{-10} \kappa^{-1} \alpha_D^{3/8} \left(\frac{v_w}{0.7}\right)^{3/2} \left(\frac{M}{10^8 {\rm g}}\right)^{-1/2} \left(\frac{\eta_\chi}{10^{-10}}\right)^{3/2} \left(\frac{\beta/H}{1000}\right)^{-3/2} \\ &\times \left(\frac{T_1}{1\textrm{ GeV}}\right)^{-3/2}  \left(\frac{m_f}{1.27\textrm{ GeV}}\right)^{-1}  \left(\ln \left(\frac{R_1}{R_{\rm tr}}\right)\right)^{1/2}~ .
\end{split}
\end{align}
Likewise, from Eqs.~\eqref{eq:evap} and~\eqref{eq:ttrhigh}, above the electroweak scale the condition is
\begin{equation}
\label{eq:formhighcond}
    \kappa < 6.70\times10^{-10} \left(\frac{M_{\rm PBH}}{10^8 ~{\rm g}}\right)^{-3/2}\left(\frac{T_1}{1\textrm{ TeV}}\right)^{-1/2}\left(\ln\Big(\frac{R_1}{R_{\rm tr}}\Big)\right)^{-1/2}~.
\end{equation}
When satisfied, the resulting PBH density at evaporation depends on the cooling time via
\begin{align}
\begin{split}
\label{eq:evapfracformhigh}
    \frac{\rho_{\rm PBH}}{\rho_{\rm SM}}\bigg|_{\rm form} =&~ 1.73\times10^{-13}\, \kappa^{-1} \alpha_D^{3/8} \left(\frac{v_w}{0.7}\right)^{3/2} \left(\frac{M}{10^8 ~{\rm g}}\right)^{-1/2} \left(\frac{\eta_\chi}{10^{-10}}\right)^{3/2} \left(\frac{\beta/H}{1000}\right)^{-3/2} \\ &\times\left(\frac{T_1}{1\textrm{ TeV}}\right)^{-1/2} \left(\ln \left(\frac{R_1}{R_{\rm tr}}\right)\right)^{1/2}~.
\end{split}
\end{align}
Note that these formulas are valid for $T_{\rm SM}^{\rm tr}<T_\star$. If $T_{\rm SM}^{\rm tr},T_{\rm SM}^{\rm evap}>T_\star$, i.e., the cooling and evaporation timescales are shorter than $t_\star$, then
the PBH energy density fraction is determined at $T_\star$. In this case, the PBH density is given by Eqs.~(\ref{eq:mass}),~(\ref{eq:nT}) as
\begin{equation}
    \frac{\rho_{\rm PBH}}{\rho_{\rm SM}}\bigg|_{\rm \star} = 2.45 \times 10^{-9}\, \alpha_D^{1/4} \left(\frac{\eta_\chi}{10^{-10}} \right)\,.
\end{equation}

\section{Regurgitated Dark Matter from Subdominant PBHs} 

PBHs that dominate the matter density can produce very heavy RDM ($\gtrsim 10^{10} \textrm{ GeV}$) or very light RDM ($\lesssim 1 \textrm{ GeV}$). PBHs that are a subdominant component can efficiently produce RDM in the WIMP mass range, $m_{(\phi,\chi)}\sim\textrm{GeV}-\textrm{TeV}$. 
We note that the condition $m_{(\phi,\chi)} > T_1, T_\star$ required for trapping
is restrictive in the case of PBH domination, but less so when PBHs are subdominant.

We estimate the PBH fraction at evaporation. Assuming no PBH domination, the Universe remains radiation dominated up to and throughout the evaporation process, with the PBHs contributing some extra radiation.
The evaporation temperature is modified from Eq.~\eqref{eq:trhm} as the PBHs no longer completely reheat the Universe:
\begin{equation}
\label{eq:evap}
    T_{\rm SM}^{\rm evap} = (43.7\textrm{ MeV})  \left(\frac{M_{\rm PBH}}{10^8~{\rm g}}\right)^{-3/2}\left(\frac{g_{\rm H, SM}}{108}\right)^{1/2}\left(\frac{g(T_{\rm SM}^{\rm evap})}{10}\right)^{-1/4}~,
\end{equation}
where $g(T_{\rm SM}^{\rm evap})$ is the number of SM d.o.f. at PBH evaporation. Assuming that the PBH lifetime is longer than the formation time, using Eqs.~\eqref{eq:mass},~\eqref{eq:nT}, and \eqref{eq:evap}, the PBH fraction at evaporation is
\begin{equation}
\label{eq:evapfrac}
    \frac{\rho_{\rm PBH}}{\rho_{\rm SM}}\bigg|_{\rm evap} = 1.43\times10^{-4}\, \alpha_D^{3/8} \left(\frac{v_w}{0.7}\right)^{3/2} \frac{M_{\rm PBH}}{10^8~{\rm g}} \left(\frac{\eta_\chi}{10^{-10}}\right)^{3/2}\left(\frac{\beta/H}{1000}\right)^{-3/2}\left(\frac{g(T_{\rm SM}^{\rm evap})}{10}\right)^{1/4}~. 
\end{equation}
The condition for PBH domination is that this quantity is larger than unity, whereas for small dark fermion asymmetries or smaller PBH masses, one has PBH evaporation in a radiation-dominated universe.
If the formation time of PBHs dominates over the evaporation time, $T_{\rm SM}^{\rm tr} < T_{\rm SM}^{\rm evap}$, a different treatment is needed. Below the electroweak scale we use Eq.~\eqref{eq:evapfracformlow}, and above the electroweak scale we use Eq.~\eqref{eq:evapfracformhigh}.
We conservatively require that PBHs both form and evaporate above $T_{\rm SM}= 5\textrm{ MeV}$.

A subdominant population of PBHs will emit a correspondingly smaller fraction of DM particles.
The density in the case of a subdominant PBH population is similar to the PBH dominant case, but with an extra factor of Eq.~\eqref{eq:evapfrac}.
For the interesting mass range from GeV to TeV, the masses of the dark sector particles are always below the Hawking temperature of the evaporating PBHs. The resulting DM density is obtained by combining of Eqs.~\eqref{eq:omegachirel} and \eqref{eq:evapfrac}:
\begin{equation}
\label{eq:omegadmsub}
    \Omega_{\rm DM (\phi,\chi)} = 3.61\times 10^{-5}\, \epsilon_{\rm em}^{-1}\, g_{H,(\phi,\chi)}\frac{m_{(\phi,\chi)}}{1\textrm{ GeV}} \left(\frac{M_{\rm PBH}}{10^8~{\rm g}}\right)^{1/2}\left(\frac{\eta_\chi}{10^{-10}}\right)^{3/2} \left(\frac{\beta/H}{1000}\right)^{-3/2}\left(\frac{g(T_{\rm SM}^{\rm evap})}{10}\right)^{-1/6}~.    
\end{equation}

\section{Interactions after the dark sector and electroweak phase transitions}
After the FOPT and electroweak symmetry breaking, the fields can be expanded around the new minima as $\phi \rightarrow \phi + v_\star$ and $\mathcal{H} = (0, \ (v_h + h))^T / \sqrt{2}$, giving the interaction Lagrangian,
\begin{eqnarray}
    \mathcal{L}_{\rm int} &=& -\frac{\kappa}{2} \phi^2 \mathcal{H}^\dagger \mathcal{H} - y_\chi \phi \Bar{\chi}\chi \nonumber \\
    &\rightarrow& - \frac{1}{2} m_{\phi,0}^2 \phi^2 - \frac{1}{2} m_{h,0}^2 h^2 - m_\chi \Bar{\chi}\chi - \frac{\kappa}{2}v_h \phi^2 h - \frac{\kappa}{4}\phi^2 h^2 - \kappa v_\star v_h \phi h - \frac{\kappa}{2} v_\star \phi h^2 - y_\chi \phi \Bar{\chi}\chi\,. \label{eq:Lrelevant} 
\end{eqnarray}
The masses include corrections arising from the vacuum expectation values, the Higgs mass includes cancellations from other contributions, and $m_\chi = y_\chi v_\star$. The fields $\phi$, $h$, and $\chi$ are in the flavor basis. 

The bilinear term in $\phi$ and $h$ implies a mixing of the form,
\begin{equation}
    \begin{pmatrix}
        \tilde{\phi} \\
        \tilde{h}
    \end{pmatrix}
    =
    U 
    \begin{pmatrix}
        \phi \\
        h
    \end{pmatrix}
    =
    \begin{pmatrix}
        \cos \theta & \sin \theta \\
        -\sin \theta & \cos \theta
    \end{pmatrix}
    \begin{pmatrix}
        \phi \\
        h
    \end{pmatrix}\,, \label{eq:mixing}
\end{equation}
where $\tilde{\phi}$ and $\tilde{h}$ are the mass eigenstates. The Lagrangian is diagonalized for the mixing angle
\begin{equation}
    \theta = \frac{1}{2} \tan^{-1} \frac{2 \kappa v_\star v_h}{m_{\phi,0}^2 - m_{h,0}^2} \,. \label{eq:theta}
\end{equation}

Hereafter, we assume that naturalness dictates $m_{\phi,0} \sim v_\star$ and $m_{h,0} \sim v_h$. Then, $\theta$ in Eq.~\eqref{eq:theta} is suppressed by the hierarchy between $v_\star$ and $v_h$. Since 
$\kappa < \sqrt{4\pi}$ by unitarity, $\theta \ll 1$ except for $m_{\phi,0} \simeq m_{h,0}$, a case that we simply discard. Then, the masses of the mass eigenstates are $m_\phi \simeq m_{\phi,0} \sim v_\star$ and $m_h \simeq m_{h,0} \sim v_h$, and the leading order Lagrangian with all self-interaction terms neglected is
\begin{eqnarray}
    \mathcal{L}'_{\rm int} &\simeq& - \frac{1}{2} m_\phi^2 \tilde{\phi}^2 - \frac{1}{2} m_h^2 \tilde{h}^2 - m_\chi \bar{\chi}\chi - y_\chi \tilde{\phi} \bar{\chi} \chi + \theta y_\chi \tilde{h} \bar{\chi} \chi \nonumber \\
    && - \frac{\kappa}{2} ( v_h + 2 \theta v_\star ) \tilde{\phi}^2\tilde{h} - \frac{\kappa}{2} ( -2 \theta v_h + v_\star ) \tilde{\phi} \tilde{h}^2 - \theta \frac{\kappa}{2} \tilde{\phi}^3 \tilde{h} - \frac{\kappa}{4} \tilde{\phi}^2 \tilde{h}^2 + \theta \frac{\kappa}{2} \tilde{\phi} \tilde{h}^3\,, \label{eq:Lpapprox}
\end{eqnarray}
where the mixing angle is
\begin{equation}
    \theta \simeq \frac{\kappa v_\star v_h}{m_\phi^2 - m_h^2}\,.
\end{equation}

We discuss several relevant interactions between the Higgs, $\phi$ and $\chi$.  In doing so, we neglect any process involving two or more particles in the initial state, assuming that they are too sparse for such an interaction to take place. 

\subsection{Decay of $\phi$}
We first calculate the decay rate of $\phi$ and check whether it can be a stable DM candidate. For $m_\phi > 2 m_h$, the leading process is the direct decay into a Higgs pair via the interaction term,
\begin{equation}
    \mathcal{L}_{\phi hh} \, \simeq \, -\frac{\kappa}{2} v_\star \tilde{\phi} \tilde{h}^2\,,
\end{equation}
giving the decay rate 
\begin{equation}
    \Gamma_{\phi \rightarrow hh} = \frac{\kappa^2 v_\star^2}{16\pi m_\phi} \left(1-4 \frac{m_h^2}{m_\phi^2} \right)^{1/2}\,.
\end{equation}
Assuming this decay channel is dominant, the lifetime of the $\phi$ particle is
\begin{equation}
    \tau_\phi = \frac{16\pi m_\phi}{\kappa^2 v_\star^2} \left(1-4 \frac{m_h^2}{m_\phi^2} \right)^{-1/2} \, \simeq \, (3.31\times 10^{-23} \, \text{sec}) \times \frac{1}{\kappa^2} \left(1-4 \frac{m_h^2}{m_\phi^2} \right)^{-1/2} \times \left(\frac{m_\phi}{1 \, \text{GeV}} \right) \left(\frac{v_\star}{1 \, \text{GeV}} \right)^{-2}.
\end{equation}
This imposes strong constraints on the Higgs portal coupling. Even the minimum requirement of $\tau_\phi \gtrsim t_U \approx 14 \, \text{Gyr}$ gives $\kappa \lesssim 10^{-22}$ for $m_\phi = 10^3$ GeV and $\kappa \lesssim 10^{-25}$ for $m_\phi = 10^9$ GeV. Such small values of $\kappa$ do not give sufficient cooling rates for $\phi$ to constitute a significant component of RDM. 

For $m_\phi \leq 2m_h$, decay into Higgs pairs is not possible and the dominant channel for $\phi$ decay is into SM fermions through their mixing with the flavor eigenstate $h$. For example, the decay rate into quarks is \cite{Gorbunov:2023lga}
\begin{equation}
    \Gamma_{\phi \rightarrow q\bar{q}} = \theta^2 \frac{N_c}{8\pi} \frac{m_q^2 m_\phi}{v_h^2} \left(1 - 4 \frac{m_q^2}{m_\phi^2} \right)^{3/2} \simeq \frac{N_c \kappa^2 v_\star^2 m_q^2 m_\phi}{8 \pi (m_\phi^2 - m_h^2)^2} \left(1 - 4 \frac{m_q^2}{m_\phi^2} \right)^{3/2} \,,\label{eq:Gammaphilight}
\end{equation}
where $N_c = 3$ is the number of colors. Assuming a single quark decay channel, the lifetime for $m_\phi^2 \ll m_h^2$ becomes
\begin{equation}
    \tau_\phi \simeq \frac{8 \pi (m_\phi^2 - m_h^2)^2}{N_c \kappa^2 v_\star^2 m_q^2 m_\phi} \left(1 - 4 \frac{m_q^2}{m_\phi^2} \right)^{-3/2} \, \sim \, (10^{-9} \, \text{sec}) \times \frac{1}{\kappa^2} \left(1 - 4 \frac{m_q^2}{m_\phi^2} \right)^{-3/2} \times \left(\frac{v_\star}{1 \, \text{GeV}} \right)^{-2} \left(\frac{m_q}{1 \, \text{MeV}} \right)^{-2} \left(\frac{m_\phi}{1 \, \text{GeV}} \right)^{-1}\,.
\end{equation}
With $m_q = 2.2$~MeV (for the up quark), $\tau_\phi \gtrsim 14 \, \text{Gyr}$ requires $\kappa \lesssim 10^{-14}$ for $m_\phi = 1 \, \text{GeV}$,  and $\kappa \lesssim 10^{-15}$ for $m_\phi = 10 \, \text{GeV}$. This again disallows $\phi$ to constitute a significant component of RDM due to a too slow cooling rate.

Thus, we conclude that $\phi$ is not a dominant RDM candidate in this particular realization. However, for the case of fermion $\chi$ RDM here, such a fast decay rate of $\phi$ simplifies the calculation and leaves $\chi$ as stable RDM.

\subsection{Invisible decay of Higgs}
We calculate the Higgs branching ratio into the dark sector. If $m_\phi < m_h / 2$, the interaction term
\begin{equation}
    \mathcal{L}_{h\phi\phi} \, \simeq \, -\frac{\kappa}{2} v_h \tilde{\phi}^2 \tilde{h}
\end{equation}
induces an invisible decay of Higgs into a pair of $\phi$'s. The corresponding decay rate is
\begin{equation}
    \Gamma_{h\rightarrow \phi \phi} = \frac{\kappa^2 v_h^2}{16\pi m_h} \left(1-4 \frac{m_\phi^2}{m_h^2} \right)^{1/2}\,.
\end{equation}
If $m_\chi < m_h / 2$,
\begin{equation}
    \mathcal{L}_{h\chi\chi} \, = \, -\theta y_\chi \tilde{h} \bar{\chi} \chi \label{eq:Lhchichi}
\end{equation}
contributes another invisible decay channel with decay rate,
\begin{equation}
    \Gamma_{h\rightarrow \chi \chi} \simeq \frac{y_\chi^2 \kappa^2 m_h}{8\pi}\left(\frac{v_\star v_h}{m_\phi^2 - m_h^2} \right)^2 \left(1 - 4\frac{m_\chi^2}{m_h^2} \right)^{3/2} = \frac{\kappa^2 v_h^2}{8\pi m_h}\left(\frac{m_\chi m_h}{m_\phi^2 - m_h^2} \right)^2 \left(1 - 4\frac{m_\chi^2}{m_h^2} \right)^{3/2}\,.
\end{equation}

The total invisible decay rate is $\Gamma_{\rm inv} = \Gamma_{h\rightarrow \phi \phi} + \Gamma_{h \rightarrow \chi \chi}$. The branching ratio is bounded by LHC data at the 95\%~C.L.~\cite{Arcadi:2019lka, Arcadi:2021mag}:
\begin{equation}
    \text{Br}(H\rightarrow \text{inv}) = \frac{\Gamma_{h \rightarrow \text{inv}}}{\Gamma_{h \rightarrow \text{inv}} + \Gamma_{h\rightarrow \text{SM}}} = \frac{\Gamma_{h \rightarrow \text{inv}}}{\Gamma_{h \rightarrow \text{inv}} + 4.07 \, \text{MeV}} < 0.11\,,
\end{equation}
which implies that 
\begin{equation}
    \Gamma_{\rm inv} = \Gamma_{h\rightarrow \phi \phi} + \Gamma_{h \rightarrow \chi \chi} < 0.50 \, \text{MeV}\,. \label{eq:Gammainvbound}
\end{equation}

Similar to $\mathcal{L}_{h\chi\chi}$, $\mathcal{L}_{\phi\chi\chi} = - y_\chi \tilde{\phi} \bar{\chi} \chi$ induces the decay of $\phi$ into $\chi$ which could increase the abundance of $\chi$ after PBH evaporation. However, since we assumed $m_\phi = m_\chi$ in Fig.~\ref{fig:exclusionsigmaDN}, the $\chi$ abundance is unaffected.

\subsection{Direct detection of $\chi$}
We obtain the direct detection bounds on $\chi$ from
$\mathcal{L}_{h\chi\chi}$, which produces inelastic scattering with nucleons, and $\mathcal{L}_{\phi\chi\chi}$, which produces elastic scattering.
By comparing the dominant latter process with the usual Higgs portal fermion DM models~\cite{Arcadi:2019lka, Arcadi:2021mag}, we get Eq.~\eqref{eq:HiggsportalsigmachiN}.

\subsection{Thermal scenarios for $\chi$}
To compute the thermal freeze-out bound for $\chi$, we assume that it is in thermal equilibrium with $\phi$ at freeze-out. Although $\chi$ can annihilate into the Higgs after both phase transitions, this process is suppressed relative to the annihilation of 
$\phi$ into the Higgs. Thus, the freeze-out of $\chi$ is determined by the freeze-out of $\phi$ if scattering and annihilation processes such as $\Gamma_{\phi \chi \xrightarrow{} \phi \chi}/H > 1$ or $\Gamma_{\phi \phi \xrightarrow{} \chi \chi}/H > 1$ are efficient at freeze-out, and keep the dark sector in equilibrium as it decouples from the SM, $\Gamma_{h h \xrightarrow{} \phi\phi}\sim 1$. This occurs if $y_\chi \gtrsim \kappa$. 

Since we assume $\phi$ and $\chi$ to be in thermal equilibrium during freeze-out, the number density of $\chi$ particles with four fermionic degrees of freedom is thrice the number density of $\phi$ with one scalar degree of freedom. So to have $\chi$ as dark matter with the correct relic energy density, the corresponding $\phi$'s abundance should be $1/3$ the $\chi$ density (assuming $m_\phi = m_\chi$), and this gives us the desired $\kappa$ for $\chi$ dark matter.

The thermal WIMPzilla line in Fig.~\ref{fig:exclusionsigmaDN} also assumes thermal equilibrium. However, this thermal WIMPzilla scenario is not a freeze-out process, but freeze-in \cite{Kolb:2017jvz}, implying that the number density of DM particles is very small after reheating and that thermal equilibrium within the dark sector is not as guaranteed as in the freeze-out case. However, since the interaction between $\phi$ and $\chi$ is generally much stronger than between SM and $\phi$, we display this freeze-in case (with internal thermal equilibrium) in the figure.

\end{document}